\documentclass[aps,prb,reprint,showpacs,longbibliography,superscriptaddress]{revtex4-2}
\usepackage{bm}
\usepackage{graphicx}
\usepackage{epstopdf}
\usepackage{latexsym}
\usepackage{subfigure}
\usepackage{color}
\usepackage{hyperref}
\usepackage{amssymb}
\usepackage{amsmath}
\usepackage{amsbsy,bm}
\usepackage{fancyhdr}
\usepackage[T1]{fontenc}
\newcommand{\CsCoBr}{{$\rm Cs_2CoBr_4$\ }}
\begin{document}

\title{Dynamics of anisotropic frustrated antiferromagnet Cs$_{2}$CoBr$_{4}$ in a spin-liquid regime}

\author{T.~A.~Soldatov}
\affiliation{P.~L.~Kapitza Institute for Physical Problems RAS, 119334 Moscow, Russia}

\author{A.~I.~Smirnov}
\affiliation{P.~L.~Kapitza Institute for Physical Problems RAS, 119334 Moscow, Russia}

\author{A. V. Syromyatnikov}
\affiliation{Petersburg Nuclear Physics Institute named by B.P. Konstantinov of National Research Center "Kurchatov Institute", Gatchina 188300, Russia}

\begin{abstract}

\CsCoBr is a triangular-lattice antiferromagnet which can be viewed as weakly interacting spin chains due to spatially anisotropic frustrating exchange couplings. The spin-orbit
interaction in Co$^{2+}$ spin-$\frac32$ ions leads to a strong easy-plane single-ion anisotropy which allows to consider the low-energy spin dynamics of this system using an
anisotropic pseudospin-$\frac12$ model. By means of the electron spin resonance (ESR) technique, we study the spin dynamics of \CsCoBr in magnetic field in a spin-liquid regime,
i.e., above the N\'eel temperature of 1.3~K but below the temperature of the crossover to in-chain correlations of pseudospins ($\approx6$~K). Our experiments reveal two bright
branches of excitations which strongly differ both from excitations in the low-temperature ordered phases and from high-temperature paramagnetic resonance of uncorrelated
pseudospins and spins.  These two branches are interpreted as excitations with zero momentum of an anisotropic spin-$\frac12$ chain. Besides, we observe several weak modes of
unknown origin which arise mostly as satellites of one of the bright modes.

\end{abstract}

\date{\today}
 \maketitle

\section{Introduction}
\label{Introduction}

\begin{figure}[t!]
\begin{center}
\vspace{0.1cm}
\includegraphics[width=0.42\textwidth]{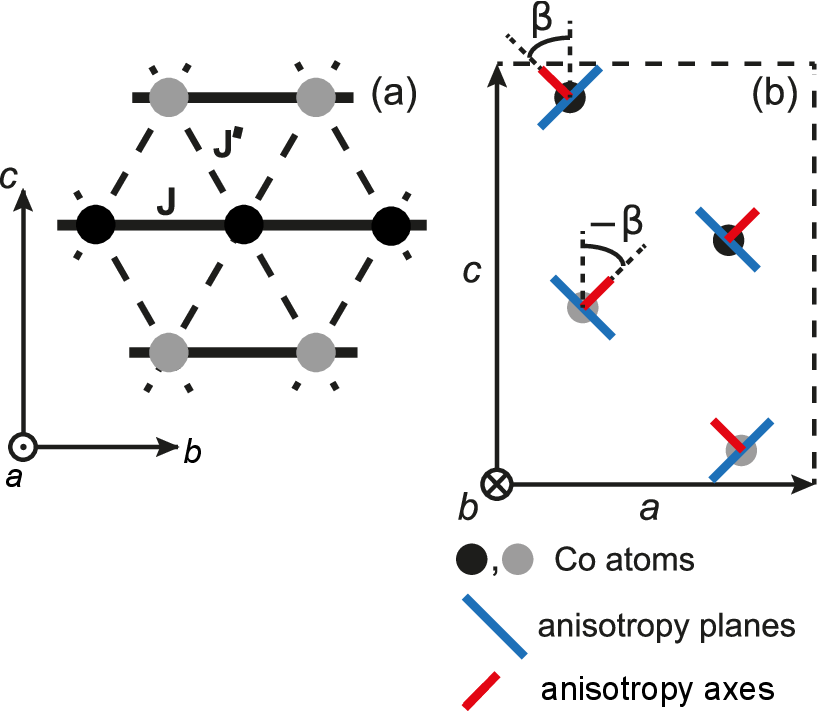}
\caption{ (a) Schematic picture of exchange paths in the $bc$ plane of Cs$_{2}$CoBr$_{4}$ and other compounds from the family $\rm Cs_2MX_4$.  (b) Simplified schematic representation of the Cs$_{2}$CoBr$_{4}$ structure projected
along the chain direction $b$. Dashed lines along $a$ and $c$ axes highlight the unit cell. Black and gray dots indicate Co atoms with crystallographic positions $y=
\frac{1}{4}b$ and $y= \frac{3}{4}b$, respectively.  Anisotropy axes and easy planes of Co$^{2+}$ ions are shown; $\beta\approx\pi/4$.\label{fig1} }
\end{center}
\end{figure}

The family of quasi-two-dimensional (quasi-2D) triangular-lattice antiferromagnets $\rm Cs_2MX_4$, where $\rm M=Cu,Co$ and $\rm X=Cl,Br$, has attracted considerable interest
recently due to rich phase diagrams in magnetic field and unusual dynamical properties
\cite{kohno,Coldea,Povarov3,Povarov1,Povarov2,Veillette,Hannahs,Smirnov,Smirnov2,Garst,Alvarez,Soldatov2023}. The combination of geometric frustration inherent to the triangular
lattice and spatial anisotropy of exchange couplings $J'<J$ shown in Fig.~\ref{fig1}(a) is responsible for the remarkable behavior of these compounds. Because of $J'<J$ and
frustration, these systems can also be viewed as weakly interacting spin chains \cite{Heidarian} passing along $b$ axis (see Fig.~\ref{fig1}).

In contrast to nearly isotropic spin-$\frac12$ Cu-based compounds of this family, Co-based materials are strongly anisotropic owing to considerable spin-orbit interaction in
spin-$\frac32$ Co$^{2+}$ ions. Single-ion anisotropy $D\approx12\,{\rm K}$ of Co$^{2+}$ ions whose easy plane alternates from chain to chain (see Fig.~\ref{fig1}) is much larger
than all exchange interactions. This allows one to describe low-energy properties of Co-based substances at $T\alt D$ by effective pseudospin-$\frac12$ anisotropic models
\cite{Povarov1,Povarov2,Povarov3,Soldatov2023,Garst,Alvarez}.

Probably the most interesting observation related to this family of materials is that spin dynamics in ordered phases of Cs$_2$CuCl$_4$ and \CsCoBr combine characteristic
features of both 1D and 2D magnets. In particular, a peculiar coexistence of two-spinon continuum of spin-$\frac12$ Heisenberg antiferromagnetic chain and quasi-2D magnons were
observed in the ordered phase of Cs$_2$CuCl$_4$  \cite{Coldea,Smirnov2,kohno}. In $\rm Cs_2CoBr_4$, numerous excitations were obtained in the ordered phase with a stripe
magnetic order by the neutron scattering and terahertz spectroscopy \cite{Povarov2,Povarov3} as well as by the ESR technique \cite{Soldatov2023}. We showed in
Ref.~\cite{Soldatov2023} that low-energy excitations are conventional spin-1 magnons and spin-0 bound states of two magnons whereas seven higher-energy modes were interpreted in
Ref.~\cite{Povarov3} as two-spinon bound states whose energies form a Zeeman ladder as in weakly coupled Ising-like spin chains.

The purpose of the present work is to study using ESR the low-energy dynamics of $\rm Cs_2CoBr_4$ in magnetic field in the spin-liquid temperature regime (i.e., above ordering
temperatures but below the characteristic energy of in-chain spin coupling). Here, we lost long range order but get a possibility to probe low energy excitations of a 1D
anisotropic quantum spin system~\cite{Coldea,Garst,Alvarez}.

We find in the present study such a regime in \CsCoBr at $1.3\,{\rm K}<T<6\,{\rm K}$. It was found in Ref.~\cite{Povarov1} that the static magnetic susceptibility deviates from the Curie-Weiss type dependence in this temperature range.
To interpret our experimental findings, we use results of numerical
investigations \cite{Garst, Alvarez} of spin-$\frac12$ XXZ chain in a transverse magnetic field $\bf h$ which were inspired by corresponding experiments in $\rm Cs_2CoCl_4$
(another member of $\rm Cs_2MX_4$ family). In contrast to this material, there are much larger in-plane inter-chain couplings and easy-axis anisotropy in $\rm Cs_2CoBr_4$. We take them into account
in a mean-field manner to reduce the model describing $\rm Cs_2CoBr_4$ to the spin-$\frac12$ XXZ chain. There is a critical field $h_c$ in the latter model separating a
low-field phase having a long-range order at $T=0$ and a collinear non-saturated state at $h>h_c$. Dynamics of the collinear phase is dominated by the low-energy gapped magnon
and the higher-energy many-particle bound state which gradually wash out at $h<h_c$ into a continuum of excitations upon the field decreasing. There are two bright bounds of
this continuum at $\bf k=0$ which we observe experimentally. We obtain also four weak ESR modes whose origin is unclear now.

We find that the paramagnetic uncorrelated regime is divided into two ranges, $6\,{\rm K}< T <15\,{\rm K}$ and $T >15\,{\rm K}$, which are governed by paramagnetic pseudospins-$\frac12$
and spins-$\frac32$, respectively. We demonstrate that the $g$-tensor which we measure in the lower paramagnetic regime differs drastically from that in the
upper regime studied in Ref.~\cite{Povarov1}. This difference is  in  a quantitative agreement with the theory.

The rest of the present paper is organized as follows. We provide details of experimental setup in Sec.~\ref{exper}. Our experimental findings are presented in
Sec.~\ref{ExpResults}. Sec.~\ref{disc} contains the discussion and theoretical interpretation of ESR spectra. An overview of results and a conclusion can be found in
Sec.~\ref{conc}.

\begin{figure}[t!]
\begin{center}
\vspace{0.1cm}
\includegraphics[width=0.42\textwidth]{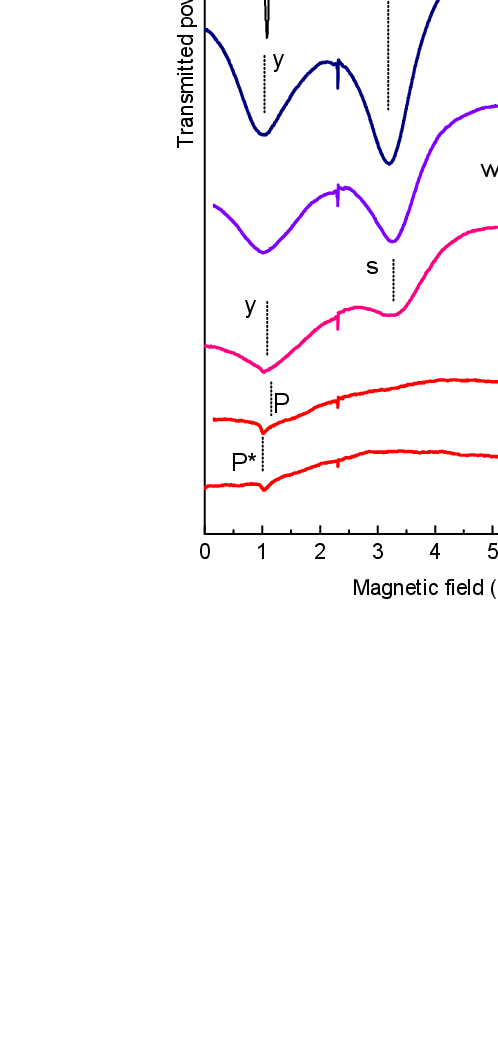}
\caption{\label{Tdep64GHz} Temperature evolution of 64.68 GHz ESR lines of \CsCoBr at ${\bf H}\parallel b$. Letters indicate modes whose frequencies are displayed on
frequency-field diagrams in Figs.~\ref{FvsHb1p5K} and \ref{FvsHb4p5K}. Vertical thick dashed lines are boundaries of ordered phases at $T=0.5$~K \cite{Povarov1,Soldatov2023}.}
\end{center}
\end{figure}

\begin{figure}[t!]
\begin{center}
\vspace{0.1cm}
\includegraphics[width=0.42\textwidth]{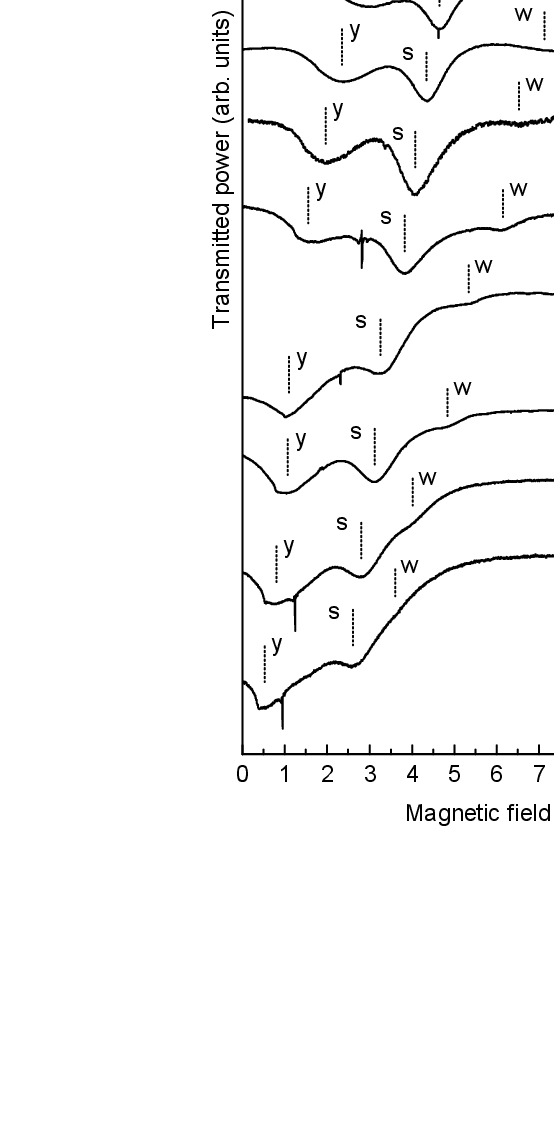}
\caption{\label{ESRlines_Hb_4p0K} ESR lines of \CsCoBr at $T=4.0$~K, ${\bf H} \parallel b$ and various frequencies. Letters indicate modes  whose frequencies are displayed on
the frequency-field diagram in Fig.~\ref{FvsHb4p5K}.}
\end{center}
\end{figure}

\section{Experiment}
\label{exper}

Experimental technique and \CsCoBr samples used are the same as described in Ref. \cite{Soldatov2023}. ESR lines have been taken at fixed frequencies from the interval
25--250~GHz as field dependencies of the microwave power transmitted through resonator containing a sample. A small amount of 2,2-diphenyl-1-picrylhydrazyl (known as DPPH) was
placed near the sample, it was used as a $g~=~2.00$ marker.   The orientation of the external field was set with the accuracy of about 2 degrees along $b$-axis. This orientation
allows to deal with the field which is within the easy planes for all four magnetic ions occupying two types of crystallographic positions in the unit cell shown in
Fig.~\ref{fig1}(b).  The easy planes of anisotropy are orthogonal for two types of nonequivalent ions, but intersect along a line which is parallel to $b$ (see Fig.~\ref{fig1}
and Refs.~\cite{Povarov1,Povarov2,Povarov3,Soldatov2023}).

\section{Experimental results}
\label{ExpResults}

As described in Ref.\cite{Soldatov2023}, we find up to seven ESR modes at a given magnetic field in ordered phases (there are also higher-energy modes above the upper limit of
our experimental setup 250~GHz which were observed in neutron and terahertz spectroscopy experiments in Refs.~\cite{Povarov2,Povarov3}). Notice that transition temperatures to
five phases at ${\bf H}\|b$ identified in Ref.~\cite{Povarov1} are smaller than $T_N=1.3$~K at $H=0$. The record at $T=0.5$~K in Fig.~\ref{Tdep64GHz} demonstrates selected
64.68~GHz ESR modes in ordered states which are marked by letters $a$, $b$, $c$, $d$, $l$, $m$, $n$, $p$, $v$, and $w$. These modes were discussed in details in
Ref.~\cite{Soldatov2023}. Other records in Fig.~\ref{Tdep64GHz} show that in the temperature range of $1.5\,{\rm K}<T<4\,{\rm K}$ the multi-mode spectrum is changed to the
spectrum consisting of two intensive lines $y$ and $s$ and a weak line $w$. As the temperature increases further, in the high-temperature range of $T>6$~K, we observe a narrow
$P^*$ and a wide $P$ ESR lines (see two bottom records in Fig.~\ref{Tdep64GHz}). Examples of ESR records taken at $T=4$~K for different frequencies in the range of 25--250~GHz
are presented in Fig.~\ref{ESRlines_Hb_4p0K}.

\begin{figure}[t!]
\begin{center}
\vspace{0.1cm}
\includegraphics[width=0.5\textwidth]{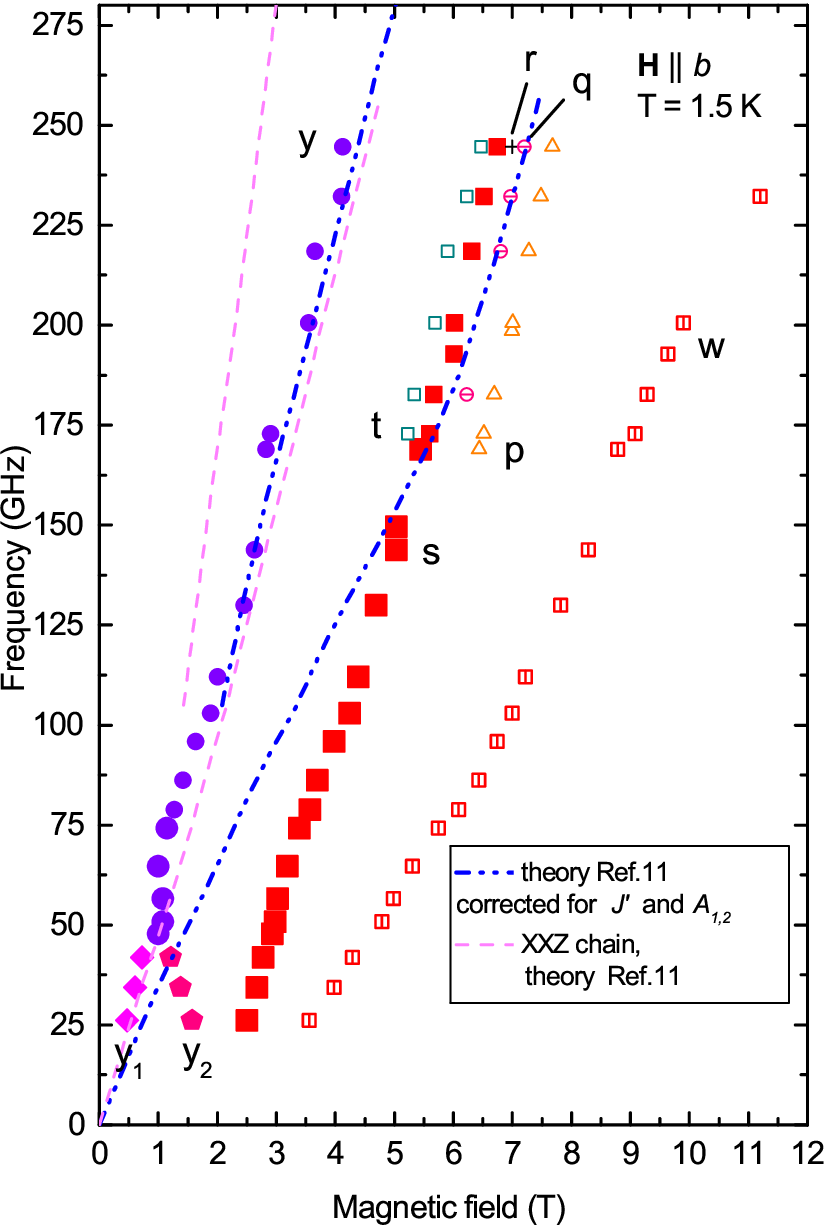}
\caption{\label{FvsHb1p5K} Frequency-field diagram of \CsCoBr at ${\bf H} \parallel b$ and $T=1.5$~K. Intensive ESR signals are marked by closed symbols whereas weak resonances
are denoted by open symbols or crosses. Dashed lines present theoretical calculations of Ref.~\cite{Alvarez} for the spin-$\frac12$ XXZ chain in a transverse field. Dash-dotted
lines show results of Ref.~\cite{Alvarez} with corrections originating from inter-chain interaction and anisotropies $A_{1,2}$ in model \eqref{hamps} (see the text). }
\end{center}
\end{figure}

\begin{figure}[t!]
\begin{center}
\vspace{0.1cm}
\includegraphics[width=0.5\textwidth]{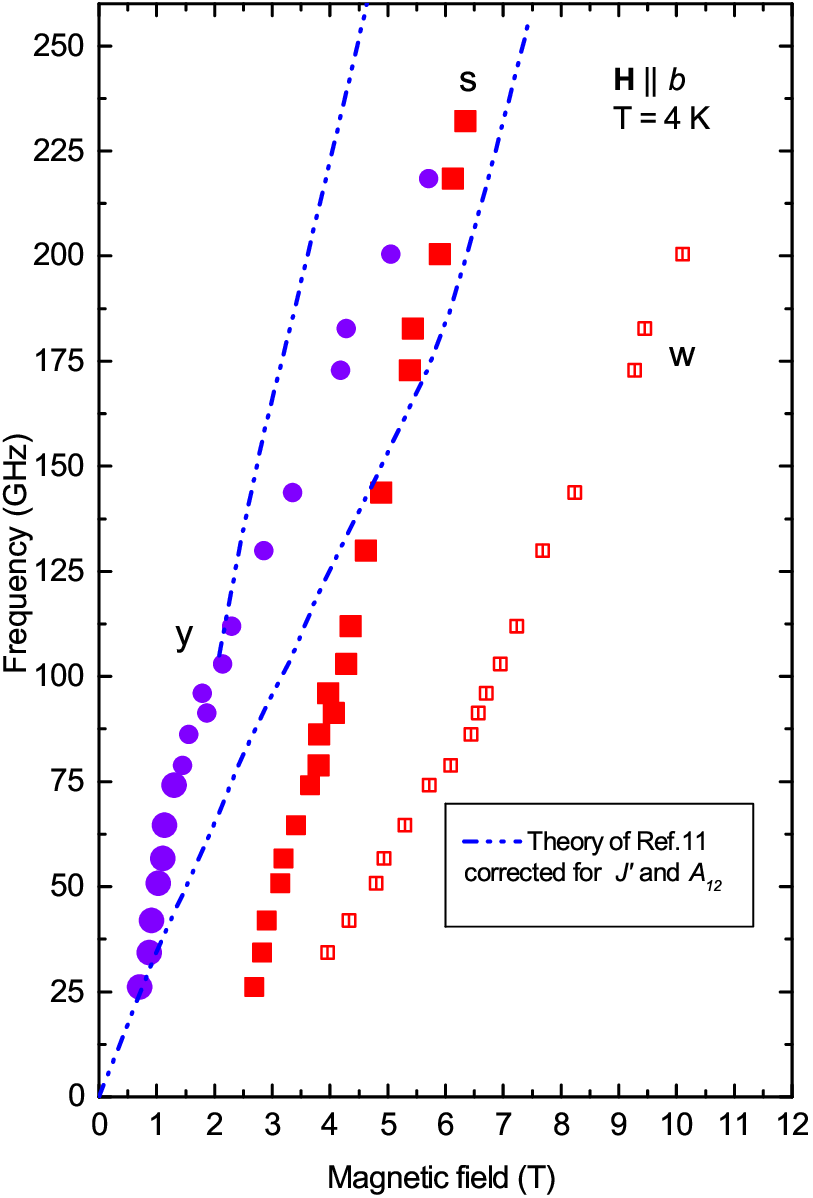}
\caption{\label{FvsHb4p5K} Same as Fig.~\ref{FvsHb1p5K} but for $T=4.0$~K.
}
\end{center}
\end{figure}

Frequency-field dependencies of ESR resonances are shown in Figs.~\ref{FvsHb1p5K} and \ref{FvsHb4p5K} for temperatures 1.5~K and 4.0~K, respectively. Data corresponding to more
intensive lines are presented by closed symbols whereas weak resonances are displayed by open symbols or crosses. As in Ref. \cite{Soldatov2023}, the bright and weak modes have
been distinguished by a rough criterion: integral intensities of "bright" and "weak" modes differ by more then three times. Resonance fields are deduced as fields of the local
maximum absorption. Because resonance lines are quite narrow, the error in the resonance field value does not exceed the size of symbols in Figs.~\ref{FvsHb1p5K} and
\ref{FvsHb4p5K}.

\begin{figure}[]
\begin{center}
\vspace{0.1cm}
\includegraphics[width=0.42\textwidth]{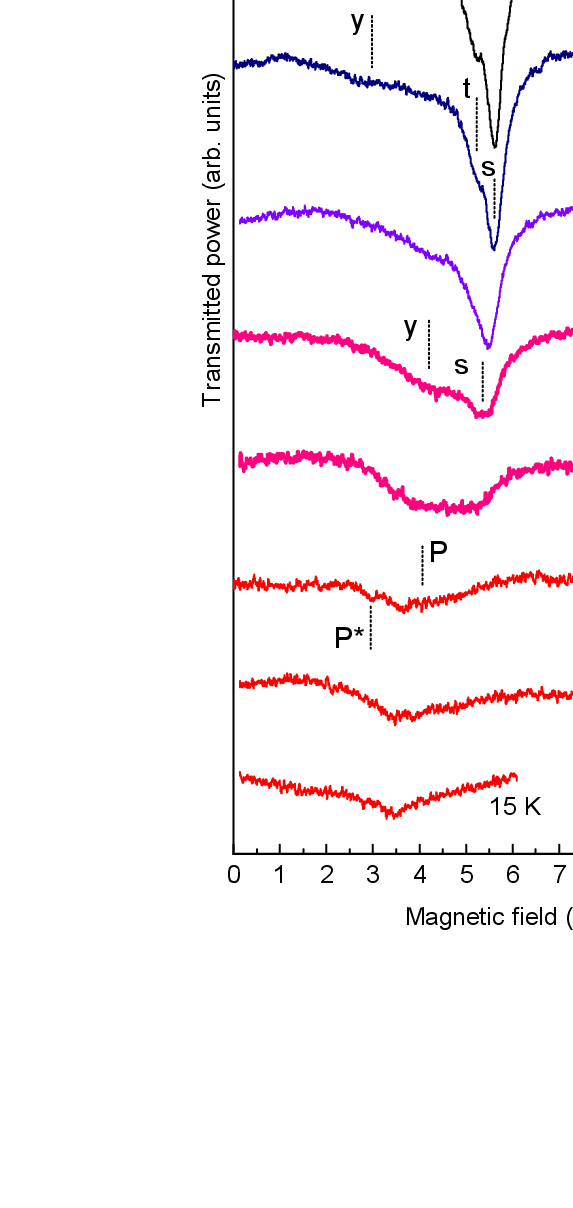}
\caption{\label{Tdep172GHz} Same as Fig.~\ref{Tdep64GHz} but for 172.86~GHz. Mode $s$ is seen to be conserved down to 0.5~K but it disappears at the crossover to the
uncorrelated state at $T\agt6$~K.
}
\end{center}
\end{figure}

Weak high-frequency resonances marked in Fig.~\ref{FvsHb1p5K} as $t$, $r$, $q$, and $p$ which look as weak satellites of the bright mode $s$ are presented in
Fig.~\ref{Tdep172GHz}. They survive down to 0.5~K and were discussed in Ref.~\cite{Soldatov2023}. Fig.~\ref{FvsHb1p5K} shows also a branching of mode $y$ into two modes $y_1$
and $y_2$ in the low-frequency range below 50~GHz. This bifurcation was also documented in Fig.~2 of our previous work \cite{Soldatov2023}.

\begin{figure}[t!]
\begin{center}
\vspace{0.1cm}
\includegraphics[width=0.42\textwidth]{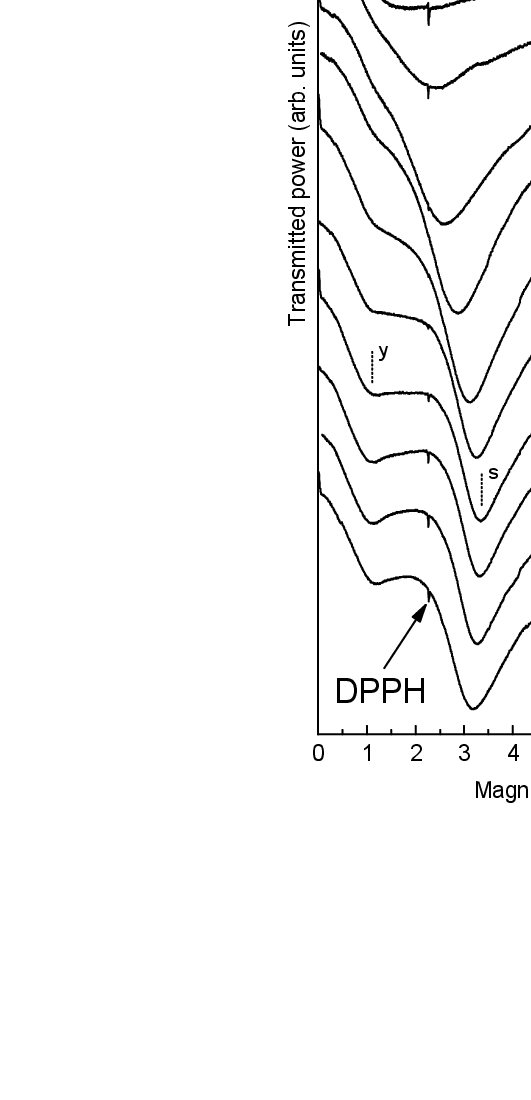}
\caption{\label{ESRlinesRota1p5K} ESR lines of \CsCoBr at 63.46~GHz, $T$=1.5~K, and at different orientation of the field in the $ab$ plane. Letters indicate modes whose
frequencies are presented in Fig.~\ref{FvsHb1p5K}. }
\end{center}
\end{figure}
\begin{figure}[t!]
\begin{center}
\vspace{0.1cm}
\includegraphics[width=0.42\textwidth]{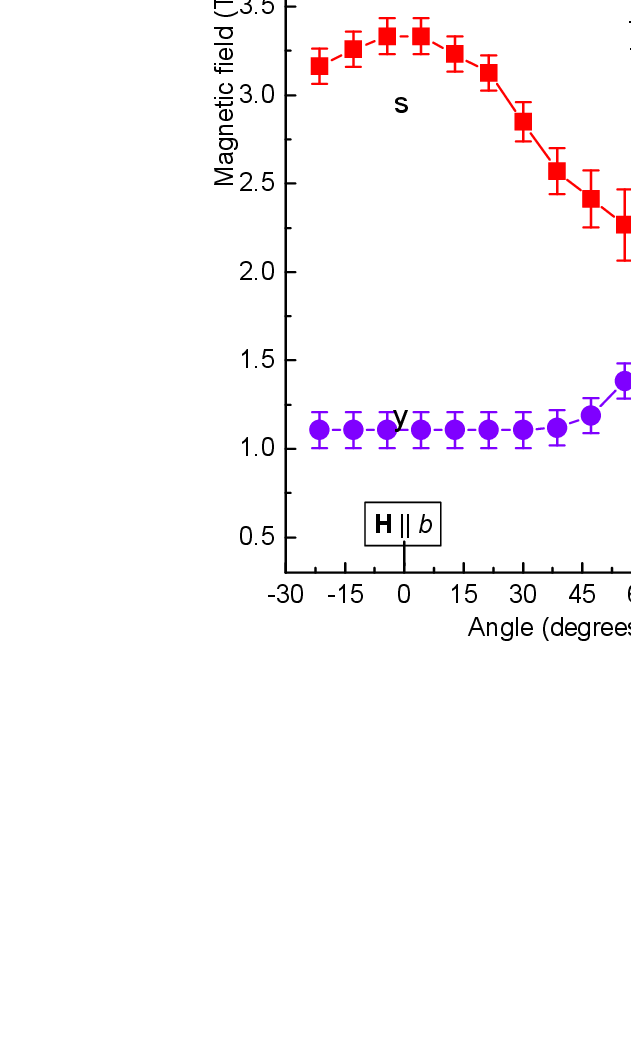}
\caption{\label{HresRota1p5K} Angular dependence of ESR signals at 63.46~GHz and $T=1.5$~K upon the field rotation in the $ab$ plane.}
\end{center}
\end{figure}

Fig.~\ref{ESRlinesRota1p5K} demonstrates the evolution of ESR records upon rotation of magnetic field in the $ab$-plane. Fig.~\ref{HresRota1p5K} shows the angular dependence
of the resonance field of the most intensive modes at $T=1.5$~K. One can see that two intensive 63.46~GHz modes observed at ${\bf H}
\parallel b$ come together and merge at ${\bf H} \parallel a$.

\begin{figure}[]
\begin{center}
\vspace{0.1cm}
\includegraphics[width=0.42\textwidth]{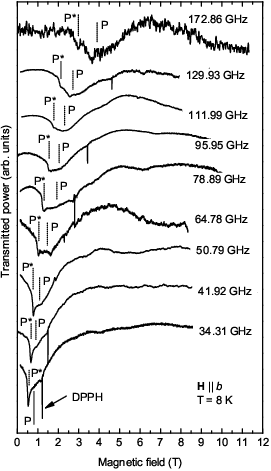}
\caption{\label{ESRlines_Hb_8p0K} ESR lines at various frequencies, ${\bf H} \parallel b$, and $T=8$~K. Resonance frequencies marked as $P$ and $P^*$ are displayed on the
frequency-field diagram in Fig.~\ref{FvsHb8p0K}. }
\end{center}
\end{figure}

\begin{figure}[]
\begin{center}
\vspace{0.1cm}
\includegraphics[width=0.42\textwidth]{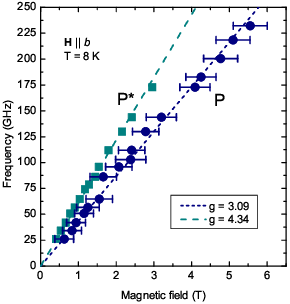}
\caption{\label{FvsHb8p0K} Frequency-field diagram of \CsCoBr at ${\bf H} \parallel b$ and $T=8.0$~K.}
\end{center}
\end{figure}

ESR lines taken at different frequencies at $T=8$~K  demonstrate the resonance absorption in the intermediate temperature range near the hump of the static susceptibility
\cite{Povarov1}, i.e., near the transition from the strongly correlated state to the regime of uncorrelated ions. These lines are  shown in Fig.~\ref{ESRlines_Hb_8p0K}. The
frequency-field diagram at $T=8$~K and ${\bf H}
\parallel b$ is presented in Fig.~\ref{FvsHb8p0K}. One can see in Figs.~\ref{Tdep172GHz}, \ref{ESRlines_Hb_8p0K}, and \ref{FvsHb8p0K} two close lines at $T>6$~K: a wide
intensive line $P$ and a weak but rather narrow line $P^*$ which have $g$-factors $g_{P} \simeq 3.1$ and $g_{P*} \simeq $4.3, respectively.  The width of line $P$ is of the
order of the resonance field. As a result, the resonance field of line $P$ has a large error of about the resonance field itself.  At higher frequencies, when the resonance
field is large, this wide resonance appears more clearly as can be seen from Fig.~\ref{ESRlines_Hb_8p0K}. Fig.~\ref{Tdep172GHz} shows that upon further temperature increasing up to 15
K the 172.86 GHz ESR line shows a single anomaly with the resonance field of 3.5 T which corresponds to a $g$-factor of 3.52. At $T=8$~K and $H\parallel c$, the $g$-factor measurement
also reveals the narrow $P^*$ and the broad $P$ lines with $g_{P^*} \simeq 4.3$ and  $g_{P} \simeq 2.8$.

\section{Discussion}
\label{disc}

To interpret our experimental findings observed above critical temperatures of transitions to ordered phases, one has to address the following spin-$\frac32$ Hamiltonian
describing $\rm Cs_2CoBr_4$ \cite{Povarov1,Povarov2,Soldatov2023}:
\begin{eqnarray}
\label{ham0} {\cal H} &=&
\sum_{i,j} \Bigl( D \left[ \left( S_{2i,j}^x \right)^2 + \left( S_{2i+1,j}^y \right)^2 \right] + J \left( {\bf S}_{i,j}{\bf S}_{i,j+1} \right)\\
&&{} + A_1 S^z_{i,j} S^z_{i,j+1} + J' \left( {\bf S}_{i,j}{\bf S}_{i+1,j} + {\bf S}_{i,j}{\bf S}_{i+1,j-1} \right)\nonumber\\
&&{} - A_2 \left( S^z_{i,j} S^z_{i+1,j} + S^z_{i,j} S^z_{i+1,j-1} \right)  \Bigr) - g\mu_BH \sum_{i,j} S^z_{i,j},\nonumber
\end{eqnarray}
where ${\bf S}_{i,j}$ is the $j$-th spin in the $i$-th chain passing along $b$ axis (see Fig.~\ref{fig1}(a)), $J>0$ and $J'>0$ are intra- and inter-chain exchange coupling
constants, respectively, $A_1\ll J$ and $A_2\ll J'$ are small anisotropies, $D\gg J,J'$ is the easy-plane anisotropy, $g\approx2.4$, a small interaction between triangular
planes and Dzyaloshinsky-Moriya interaction (DMI) inherent to $\rm Cs_2MX_4$ compounds are omitted, $z$ and $b$ axes are parallel to each other, $x$ and $y$ axes are mutually
orthogonal and they are parallel to hard axes in neighboring chains shown in Fig.~\ref{fig1}(b).

Our previous consideration of spin dynamics in \CsCoBr using the bond-operator-theory \cite{Soldatov2023} provides the following set of model parameters which describe well ESR
and neutron spectra in the ordered phases:
\begin{equation}
\label{param}
\begin{aligned}
    J &= 0.165~{\rm meV}, & A_1 &= 0.34J,\\
    J' &= 0.45J, &  A_2 &= 0.1J'.
\end{aligned}
\end{equation}
The single-ion anisotropy was estimated in Ref.~\cite{Povarov1} as $D\approx12\,{\rm K}\approx1\,{\rm meV}$.

In the local coordinate frame in which the quantized axis is directed perpendicular to the easy plane, the lower doublet |$\pm$1/2$\rangle$ of each spin is separated by $2D$
from the upper doublet |$\pm$3/2$\rangle$. Due to large $D\gg J,J'$, this allows to discuss the low-energy dynamics and low-temperature behavior (at $T\alt D$) on a simpler
model by introducing pseudospin-$\frac12$ ${\bf s}_{i,j}$ at each lattice site which describes the lower doublet |$\pm$1/2$\rangle$. The transition from spins to pseudospins can
be made according to the rule which readily follows from the comparison of matrix elements of spin operators 1/2 and 3/2:
\begin{equation}
\label{rule}
\begin{aligned}
S^x_{2i,j}&\mapsto s^x_{2i,j}, & S^y_{2i,j}&\mapsto 2s^y_{2i,j}, &
S^z_{2i,j}&\mapsto 2s^z_{2i,j}, \\
S^x_{2i+1,j}&\mapsto 2s^x_{2i+1,j}, & S^y_{2i+1,j}&\mapsto s^y_{2i+1,j}, & S^z_{2i+1,j}&\mapsto 2s^z_{2i+1,j}.
\end{aligned}
\end{equation}
As a result, one comes from Eq.~\eqref{ham0} to the following pseudospin-$\frac12$ Hamiltonian: \cite{Povarov1,Soldatov2023}
\begin{eqnarray}
\label{hamps} {\cal H}_{ps} &=& \sum_{i,j} \left( 4J \left( {\bf s}_{2i,j}{\bf s}_{2i,j+1} \right) - 3J s^x_{2i,j} s^x_{2i,j+1}
\right.\\
&&{}
 + \left. 4J \left( {\bf s}_{2i+1,j}{\bf s}_{2i+1,j+1} \right) - 3J s^y_{2i+1,j} s^y_{2i+1,j+1}
\right.\nonumber\\
&&{} + \left. 4A_1 s^z_{i,j} s^z_{i,j+1} + 2J' \left( {\bf s}_{i,j}{\bf s}_{i+1,j} \right)
\right.\nonumber\\
&&{} + \left. 2J' \left( {\bf s}_{i,j}{\bf s}_{i+1,j-1} \right) + \left( 2J' - 4A_2 \right) s^z_{i,j} s^z_{i+1,j}
\right.\nonumber\\
&&{} \left. + \left( 2J' - 4A_2 \right) s^z_{i,j} s^z_{i+1,j-1} \right) - 2g\mu_BH \sum_{i,j} s^z_{i,j}.\nonumber
\end{eqnarray}

Ordered phases of model \eqref{hamps} arise at $T<1.3$~K due to inter-chain coupling $J'$ and omitted small inter-plane interaction. The spin-liquid regime in which properties
of the system are governed by in-chain spin interactions is expected at $1.3\,{\rm K}\alt T\alt 4 (J+A_1)\approx10$~K. Notice that the upper limit of this estimation fulfills
the criterion of applicability of the pseudospin treatment $T\ll D$. Simple putting $J'=A_{1,2}=0$ in Eq.~(\ref{hamps}) reduces the model to the spin-$\frac12$ XXZ chain in a
transverse magnetic field considered numerically, in particular, in Ref.~\cite{Alvarez} in relation with experiments on $\rm Cs_2CoCl_4$ in the spin-liquid regime. Notice that
the description of $\rm Cs_2CoCl_4$ at $T>T_N$ by the spin-$\frac12$ XXZ chain is quite adequate due to a negligible $J'\ll J$. In contrast, terms $J'$ and $A_1$ in model
(\ref{hamps}) which differ \CsCoBr from an array of XXZ chains cannot be simply discarded in a quantitative consideration (see Eq.~\eqref{param}). Nevertheless, we show below
that numerical considerations of the spin-$\frac12$ XXZ chain performed in Refs.~\cite{Garst,Alvarez} can be adopted for a quantitative description of our results in \CsCoBr by
taking into account terms $J'$ and $A_{1,2}$ in a simple mean-field manner.

Then, let us discuss the spin-$\frac12$ XXZ chain in a transverse field described by the Hamiltonian
\begin{equation}
\label{xxz} {\cal H}_c = \sum_i\left[ J_c \left( \Delta s^x_is^x_{i+1} + s^y_is^y_{i+1} + s_i^zs_{i+1}^z \right) - hs_i^z \right].
\end{equation}
In the case of $-1<\Delta<1$ which is relevant to our consideration, there are two phase transitions in model \eqref{xxz}. The first one takes place at $h=0$, where the system
is equivalent to a gapless Luttinger liquid \cite{vidal}. The second one, occurs at $h=h_c>0$ between a spin-flop gapped phase with a long-range order at $0<h<h_c$ (analogous to
that in a 3D antiferromagnet in a field above the spin-flop transition) and a spin-polarized non-saturated state at $h>h_c$. It was found in Ref.~\cite{Alvarez} that
$h_c\approx1.6 J_c$ at $\Delta=0.25$. The spectrum of spin excitations has a continuum-like character at $h<h_c$ with two quite bright boundaries (at $h=0$, the spectrum is
dominated by a two-spinon continuum) which transform upon approaching $h_c$ into two sharp coherent modes which are interpreted at $h>h_c$ as a low-energy gapped magnon mode and
a high-energy many-particle bound state  \cite{Alvarez,Garst}. The temperature impact on spectra becomes noticeable starting from $T\approx J_c/4$  \cite{Garst}.

Let us treat terms $J'$ and $A_{1,2}$ in model \eqref{hamps} in a mean-field-like fashion by replacing in them ${\bf s}_{i,j}{\bf s}_{k,q}\mapsto (s^z_{i,j} + s^z_{k,q})\langle
s^z\rangle$ and $s^z_{i,j}s^z_{k,q}\mapsto (s^z_{i,j} + s^z_{k,q})\langle s^z\rangle$, where $\langle s^z\rangle$ is the uniform longitudinal pseudospin magnetization and we
assume also that thermal fluctuations melt the long-range order at temperatures of our current interest. As a result, we come from Eq.~\eqref{hamps} to Eq.~\eqref{xxz}, where
\begin{eqnarray}
\label{corr}
    J_c &=& 4J,\nonumber\\
    \Delta &=& 1/4,\\
    h &=& 2g\mu_BH-8(A_1+2(J'-A_2)) \langle s^z\rangle. \nonumber
\end{eqnarray}
We can adopt now results of Ref.~\cite{Alvarez}, where model \eqref{xxz} was considered with $\Delta=0.25$. Notice also that excitation spectra found in Ref.~\cite{Alvarez} do
not differ much from those in Ref.~\cite{Garst}, where $\Delta=0.12$ was discussed.
For comparison of our ESR data with the single-chain theory, we take energies of anomalies in the dynamical structure factor at $k=0$ from Fig.~S12 of the Supplemental materials
of Ref.~\cite{Alvarez} and recalculate them and corresponding field values using Eqs.~\eqref{param} and \eqref{corr}. The pseudospin magnetization $\langle s^z\rangle$ is taken
from Fig.~6 of Ref.~\cite{Povarov1} in which the experimental magnetization curve in \CsCoBr at $T=1.8$~K is presented (we take into account also that according to
Eq.~\eqref{rule} the mean spin magnetization $\langle S^z\rangle$ is related to the pseudospin one as $\langle S^z\rangle=2\langle s^z\rangle$).

Numerical data recalculated in this way are presented by dash-dotted lines in Figs.~\ref{FvsHb1p5K} and \ref{FvsHb4p5K}. Notice that according to Eq.~\eqref{corr}
$h_c\approx1.6J_c$ corresponds to $H_c\approx6.3$~T. One can see from Fig.~\ref{FvsHb1p5K} that the most intensive ESR signals $s$ and $t$ are described quantitatively at
$T=1.5\,{\rm K}\approx 0.2J_c$ and $H\approx H_c$ by spectra of two coherent modes discussed in Refs.~\cite{Alvarez,Garst}. Fig.~\ref{FvsHb4p5K} demonstrates that at larger
temperature $T=4\,{\rm K}\approx 0.5J_c$ the upper ESR signal $t$ tends to merge with lower $s$ mode at $H\approx H_c$ that is in a qualitative agreement with results of
numerical investigation of thermal effect on spectra (see Fig.~5 in Ref.~\cite{Garst}). It was speculated in Ref.~\cite{Garst} that thermal fluctuations destroy many-particle
bound states at $T\sim J_c$ and produce an anomaly near $\bf k=0$ just above the lower mode which is associated with the bound states decay. At $H<4$~T, the ESR mode $s$
deviates from the theoretical curve in Figs.~\ref{FvsHb1p5K} and \ref{FvsHb4p5K} that may be an indication that our mean-field treatment is too rough of $J'$ and $A_{1,2}$ terms
in Hamiltonian \eqref{hamps}.

Notice that both Co- and Cu-based members of the family $\rm Cs_2MX_4$ obey continuums of excitations at not too large fields despite they are described by strongly anisotropic
model \eqref{ham0} and an isotropic Heisenberg models (with only small anisotropic terms), respectively  \cite{Smirnov,Alvarez,Garst,Soldatov2023}. Let us compare now these
continuums and discuss how they appear in experiments.
In the two-spinon continuum obtained analytically in a free-fermion approximation in the spin-$\frac12$ Heisenberg chain \cite{Dender}, the excitation frequency at $\bf k=0$
(which is relevant to the ESR experiment) is equal to the Larmor frequency. At this frequency, the upper and the lower boundaries of the so-called two-spinon continuum merge.
Nevertheless the numerical method of Ref.~\cite{kohnoPRL} as well as the analytical approach of Refs.~\cite{Starykh2020,Povarov2022} show that there is a gap between the Larmor
mode and the continuum at $\bf k=0$ shifted upwards in energy (with a zero dynamical susceptibility for the continuum at $\bf k=0$). It was shown in Ref.~\cite{Povarov2022} that
this gap is caused by the interaction of spinons. In the presence of the low-symmetry uniform DMI, this susceptibility becomes nonzero and was observed in the ESR experiment
\cite{Povarov2022}. Two obtained ESR lines marking the lower and the upper boundaries of the continuum were named "spinon doublet" in Heisenberg spin-$\frac12$ chain compounds
with the uniform DMI. The gap between components of the doublet depends on the DMI value and on the field-induced interaction term \cite{Povarov2022}.  This doublet was
observed, e.g., in Cs$_2$CuCl$_4$ \cite{Smirnov}, K$_2$CuSO$_4$Br$_2$ \cite{Smirnov2015,Povarov2022}, Na$_2$CuSO$_4$Br$_2$ \cite{Ohta}, and Ca$_3$ReO$_5$Cl$_2$ \cite{zvyag}.

Numerical results for the spin-$\frac12$ XXZ chain in a transverse field \cite{Garst, Alvarez} also show a continuum with a field-induced gap between the upper and the lower
boundaries at $\bf k=0$ which is very similar to that of the Heisenberg model \cite{kohnoPRL}. The frequency of the lower $\bf k=0$ excitation in Figs.~\ref{FvsHb1p5K} and
\ref{FvsHb4p5K} is near the pseudospin Larmor frequency $2g\mu_BH/\hbar$  and the dynamical susceptibility of the upper branch is also predicted to vanish at $\bf k=0$. We
propose that the symmetry-allowed uniform DMI \cite{Smirnov} and the anisotropy in the $(ac)$-plane lead to the interaction between the microwave magnetic field and spin
oscillations that allows to observe in \CsCoBr both resonance frequencies predicted by the XXZ-chain-theory \cite{Garst,Alvarez}.



The anisotropic behavior of two lines $s$ and $y$, presented by the angular dependence of the resonance fields in Fig.~\ref{HresRota1p5K} does not match the angular dependence
of the spinon doublet in Cs$_2$CuCl$_4$, where the doublet lines were separated at ${\bf H}\parallel a$ and merged at ${\bf H}\parallel b$. This probably means that the angular
dependence in \CsCoBr is mainly provided by the anisotropy rather than by the DMI.

Noteworthy, the uncorrelated paramagnetic region has its own structure in $\rm Cs_2CoBr_4$.
  As mentioned in Sec. \ref{ExpResults},
at $6\,{\rm K}<T<15\,{\rm K}\sim2D$, the observed effective $g$-factor of the narrow line $P^*$ is 4.3 for ${\bf H} \parallel b$ (see Fig.~\ref{FvsHb8p0K}) and  at ${\bf H}
\parallel c$. On the other hand, $g\approx g_0=2.4$ was obtained in Ref.~\cite{Povarov1} at $T>15$~K by a mean-field analysis of static susceptibilities. The origin of this
discrepancy is that the pseudospin concept remains valid at $6\,{\rm K}<T<15\,{\rm K}$ although the in-chain spin correlations are destroyed at such $T$. As a result, the Zeeman
interaction is given by the last term in pseudospin Hamiltonian \eqref{hamps} in which $g$-factor is effectively doubled as a result of the transition from spins to pseudospins.
Then, the effective $g$-factor should equal $2g_0\approx4.8$ at $6\,{\rm K}<T<15\,{\rm K}$ for ${\bf H } \parallel b$ in a good agreement with our data. According to
Eqs.~\eqref{rule}, $g$-factor should be $g_0\sqrt{5/2}\approx3.8$ at ${\bf H}\parallel c$ in this temperature region that also approximately agrees with our findings.

The wide line $P$ with $g_P \approx 3$ comes at $6\,{\rm K}<T<15\,{\rm K}$ probably from the spin-liquid resonances $s$ and $y$ smeared by thermal fluctuations. As it is seen from the ESR record at $T=15\,{\rm K}$ presented in Fig.~\ref{ESRlines_Hb_8p0K}, ESR lines with modes $P$ and $P^*$ transform into lines with a single mode obeying the intermediate $g$-factor of
about 3.5 upon the temperature rising. The value of $g\approx2.4$ obtained from the high-temperature susceptibility fit in
Ref.~\cite{Povarov1} indicates that an additional transformation of the ESR spectrum is expected at $T>15\,{\rm K}$ which is caused by the transition from pseudospins-$\frac12$ to
spins-$\frac32$. We do not study this transition experimentally in the present work.

\section{Summary and conclusion}
\label{conc}

To conclude, we perform ESR investigation of quasi-2D triangular-lattice \CsCoBr in magnetic field $\bf H$ and reveal a drastic change of spin dynamics with temperature
increase.

We observe up to nine modes in field-induced ordered phases at $T<1.3\,{\rm K}$ which we discussed in detail in Ref.~\cite{Soldatov2023}. This rich dynamics is caused by strong
spatial anisotropy of exchange couplings and single-ion easy-plane anisotropy $D\approx12\,{\rm K}$ of spin-$\frac32$ Co$^{2+}$ ions due to which the system can be viewed as an
array of weakly coupled anisotropic chains passing along $b$ axis (see Fig.~\ref{fig1}). Large $D$ exceeding well values of all other spin interactions allows to describe
low-energy dynamics at $T\ll 2D$ by effective pseudospin-$\frac12$ model \eqref{hamps}. In particular, we showed in Ref.~\cite{Soldatov2023} that low-energy excitations are
conventional spin-1 magnons and spin-0 bound states of two magnons in the stripe and in the "up-up-down" states. The set of higher-energy modes in the stripe phase were
interpreted in Ref.~\cite{Povarov3} as a Zeeman ladder of two-spinon bound states characteristic of strongly anisotropic spin chains.

We find in the present study a spin-liquid regime in \CsCoBr at $1.3\,{\rm K}<T<6\,{\rm K}$ in which dynamics is governed by in-chain spin correlations. To interpret our
experimental findings, we use results of numerical investigations \cite{Garst, Alvarez} of spin-$\frac12$ XXZ chain in a transverse magnetic field which were inspired by
corresponding experiments in isostructural compound $\rm Cs_2CoCl_4$. In contrast to this material, there are much larger in-plane inter-chain coupling and easy-axis anisotropy
in \CsCoBr which we take into account in a mean-field fashion to reduce our model to the spin-$\frac12$ XXZ chain. There is a critical field $h_c$ in the latter model separating
a low-field phase having a long-range order at $T=0$ and a collinear non-saturated state at $h>h_c$. Dynamics of the latter phase is dominated by the low-energy gapped magnon
and the higher-energy many-particle bound state which gradually wash out at $h<h_c$ into a continuum of excitations upon the field decreasing. There are two bright bounds of
this continuum at $\bf k=0$ which can be seen in our ESR experiment. Fig.~\ref{FvsHb1p5K} shows that our two most bright ESR signals $y$ and $s$ are described quantitatively by
spectra of these coherent modes at $h\approx h_c$ and $T=1.5$~K ($h_c$ corresponds to the field of 6.3~T in $\rm Cs_2CoBr_4$). At $h\ll h_c$, the agreement is much worse
presumably due to our rough mean-field treatment of the inter-chain interaction and the anisotropy. The origin of weak ESR modes $t$, $r$, $q$, and $p$ in Fig.~\ref{FvsHb1p5K},
which are satellites of the bright mode $s$, remains unexplained that leaves room for further theoretical efforts in this field. The merging of $y$ and $s$ lines in
Fig.~\ref{FvsHb4p5K} at $T=4$~K is qualitatively attributed to the temperature effect discussed in Ref.~\cite{Garst}.

We provide the evidence of the internal structure of the uncorrelated paramagnetic regime at $T>6\,{\rm K}$. It arises due to the temperature transition
 from pseudospin-$\frac12$ to spin-$\frac32$ treatments of magnetic Co$^{2+}$ ions which are valid at $T\alt15\,{\rm K}\sim 2D$ and $T\agt15\,{\rm K}$, correspondingly.
  At $6\,{\rm K}<T<10\,{\rm K}$, we observe the sharp $P^*$ and the wide $P$ ESR signals attributed, respectively, to isolated pseudospins and remnants of coherent
  modes $s$ and $y$ washed out by thermal fluctuations.
In agreement with the theory, our ESR data show that the $g$-factor of the sharp $P^*$ mode is about two times larger at ${\bf H}\parallel b$ than
the $g$-factor deduced in Ref.~\cite{Povarov1} from the analysis of the static susceptibility at $T>15\,{\rm K}$.


\begin{acknowledgments}

We acknowledge K.\ Yu.~Povarov and A.~Zheludev for presenting the samples for this investigation and discussions, S.S. Sosin for numerous discussions. A.\ V.~Syromyatnikov
acknowledges the financial support by the Russian Science Foundation (Grant No.\ 22-22-00028). Work on ESR experiment in Kapitza Institute was supported by Russian Science
Foundation (Grant No.\ 22-12-00259).

\end{acknowledgments}

\bibliography{lit4CCBv20Aug23}

\end{document}